# Space–Time Parity Violation and Magnetoelectric Interactions in Antiferromagnets

A. M. Kadomtseva[1], A. K. Zvezdin[2], Yu. F. Popov[1,]*, A. P. Pyatakov[1], and G. P. Vorob'ev[1]

[1] *Moscow State University, Vorob'evy gory, Moscow, 119992 Russia*
*e-mail: Popov@plms.phys.msu.ru*
[2] *Institute of General Physics, Russian Academy of Sciences, ul. Vavilova 38, Moscow, 119991 Russia*

The properties of antiferromagnetic materials with violated space–time parity are considered. Particular attention is given to the bismuth ferrite $BiFeO_3$ ferroelectric magnet. This material is distinguished from other antiferromagnets in that the inversion center is absent in its crystal and magnetic structures. This circumstance gives rise to the diversified and unusual properties, namely, to the appearance of a spatially modulated spin structure and to the unique possibility of the linear magnetoelectric effect coexisting with a weak ferromagnetic moment. The magnetic-induced phase transitions accompanied by the suppression of the modulated spin structure and appearance of a number of new and unusual effects are considered. These are the linear magnetoelectric effect and the appearance of a toroidal moment and a weak ferromagnetic moment of the magnetoelectric nature.

## I. INTRODUCTION

The magnetic symmetry method is an elegant and efficient tool for studying the physical properties of crystals with complex magnetic structure, notably antiferromagnets. On the basis of this approach, the Dzyaloshinski–Moriya interaction and weak ferromagnetism [1–3], the linear magnetoelectric effect [4–6], piezomagnetism [7, 8], linear magnetostriction [9], and a variety of unusual optical effects associated with the antiferromagnetic vector, e.g., quadratic Faraday effect [10] or linear birefringence [11, 12], were discovered in the past century. Similar effects also occur in kinetics [13] (see also [14–16]).

The occurrence of one or other of these effects in an antiferromagnetic crystal with the centrosymmetric crystallographic structure depends on the space parity $I$ of its magnetic structure. For example, weak ferromagnetism arises in crystals with the even magnetic structure, whereas the linear magnetoelectric effect is forbidden in them. Quite the reverse, the linear magnetoelectric effect is allowed and a weak ferromagnetism is forbidden in crystals with the odd magnetic structure. In this respect, crystals with trigonal symmetry—rhombohedral $MnCo_3$, $FeBO_3$, and $\alpha$-$Fe_2O_3$ antiferromagnets with the even magnetic structure and $Cr_2O_3$ antiferromagnet with the odd magnetic structure—have received the most study. The crystal structures of all these materials possess the inversion center (they belong to the space group $R3\bar{c} = D_{3d}^6$). In this work, the physical properties of bismuth ferrite $BiFeO_3$ (and other materials on its base) are considered. This material is related to the aforementioned rhombohedral antiferromagnets but is different from them in that the inversion center is absent in both its crystal and magnetic structures. It will be shown that quite diversified and uncommon properties of this material are due precisely to this fact.

## II. CRYSTAL AND MAGNETIC STRUCTURES OF $BiFeO_3$

The crystal structure of bismuth ferrite is characterized by a rhombohedrally distorted perovskite unit cell (Fig. 1a) that is very close to a cube with the edge $a_c = 3.96$ Å and the almost direct angle $\alpha = 89.4°$ at its face. However, when analyzing the properties of bismuth ferrite, it is more suitable to consider a hexagonal cell (Fig. 1b) with the parameters $a_{hex} = 5.58$ Å and $c_{hex} = 13.9$ Å. The iron and bismuth ions are offset from their centrosymmetric positions, resulting in the spontaneous polarization $P_s$ along the $[111]_c$ ($[001]_{hex}$) direction [17]. Early neutron diffraction studies [18] have shown that bismuth ferrite possesses the antiferromagnetic ordering of the $G$ type, where every atom is surrounded by six atoms with the oppositely oriented spins. This signifies that bismuth ferrite is a ferroelectric magnet [19], i.e., a material with the coexisting magnetic and electric order parameters.

More precise measurements performed on a time-of-flight neutron diffractometer [20] revealed a more complicated spatially modulated magnetic structure with a large period $\lambda = (620 \pm 20$ Å$)$ incommensurate with the lattice parameter. The magnetic moments of iron ions retain their local mutually antiferromagnetic G-type orientation and rotate along the propagation direction of the modulated wave in the plane perpendicular to the hexagonal basal plane. The spin distribution along the spatially modulated structure was determined from the NMR spectra obtained in the experiments of Zalesskiĭ et al. in [21–23].

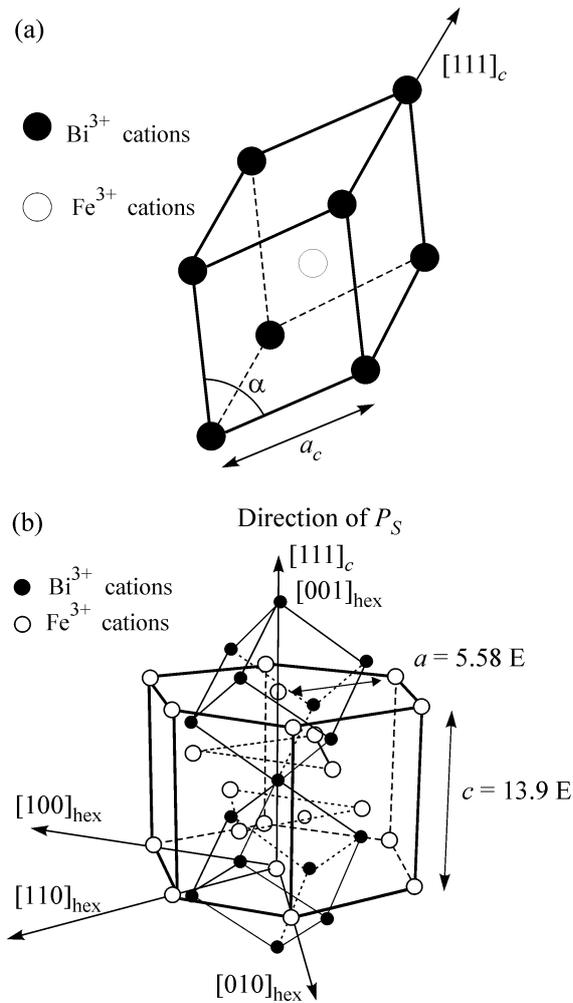

**Fig. 1.** Perovskite crystal structure: (a) rhombohedral and (b) hexagonal unit cells.

### III. SYMMETRY AND MAGNETOELECTRIC INTERACTIONS

To analyze the magnetoelectric properties of bismuth ferrite, we choose the space group $R\bar{3}c$ as a "parent" symmetry for the $R3c$ symmetry under study. The latter differs from $R\bar{3}c$ only by the presence of a polar vector $\mathbf{P} = (0, 0, P_s)$. Although, in reality, the phase transition in BiFeO$_3$ at the Curie point $T_C$ differs from the assumed $R3c \longrightarrow R\bar{3}c$ transition, this is of no importance for our purposes, i.e., for determining the adequate invariants that are responsible for the properties of the system. Indeed, the parent symmetry $R\bar{3}c$ can be used to develop the perturbation theory for determining the thermodynamic potential and other physical quantities of the system at the sacrifice of the smallness of $\mathbf{P}$; i.e., we develop the perturbation theory with respect to $\mathbf{P}$. Correspondingly, $\xi = \Delta a/a$ is a small parameter, where $a$ is the lattice constant, and $\Delta a$ is the characteristic atomic deviation from the positions symmetric about the space inversion in $R3c$.

The exchange magnetic structure, i.e., the mutual directions of magnetic moments in crystal, is determined by the following code (Turov nomenclature [24]): $I^-$, $3_Z^+$, $2_X^+$, where the space-inversion element $I$, the threefold axis $3_Z$ (aligned with the $c$ axis), and the twofold axis $x$ are the group generators and the indices $\pm$ of these elements specify their parity about the transposition of magnetic sublattices; i.e., the sign "+" indicates that the symmetry element transposes the ions within the same magnetic sublattice of an antiferromagnet, and the sign "–" indicates that the sublattice is transposed into the sublattice with the opposite spin direction upon the symmetry operation. With these symmetry operators, the antiferromagnetic vector obeys the following transformation rules:

$$I^\pm \mathbf{L} = \pm \mathbf{L}, \quad 2_X^\pm L_X = \pm L_X, \quad 2_X^\pm L_{Y(Z)} = \mp L_{Y(Z)}.$$

For the other vectors, the action of elements with different indices is the same:

$$I^\pm \mathbf{m} = \mathbf{m}, \quad 2_X^\pm m_X = m_X, \quad 2_Z^\pm m_{Y(Z)} = -m_{Y(Z)},$$

$$I^\pm \mathbf{P} = \mathbf{P}, \quad 2_X^\pm P_X = P_X, \quad 2_X^\pm P_{Y(Z)} = -P_{Y(Z)}.$$

The layout of the group elements relative to the magnetic ions in bismuth ferrite is shown in Fig. 2a. Of interest is to compare the code of this structure (BiFeO$_3$) with the codes of other antiferromagnetic compounds belonging to the space group $R\bar{3}c$: $I^+$, $3_X^+$, $2_X^+$ for hematite ($\alpha$-Fe$_2$O$_3$) and $I^-$, $3_Z^+$, $2_X^-$ for chromite (Cr$_2$O$_3$), whose exchange structures are shown in Figs. 2b and 2c, respectively. As far as we know, materials with the $I^-$, $3_Z^+$, $2_X^+$ code have not been considered as yet.

The space group $R\bar{3}c$ contains eight irreducible representations: four one-dimensional ($\Gamma_1$, $\Gamma_2$, $\Gamma_4$, $\Gamma_5$) and four two-dimensional ($\Gamma_3$, $\Gamma_3'$, $\Gamma_6$, $\Gamma_6'$) representations (Table 1). Their matrix representations are given in the columns corresponding to the generating symmetry elements. The vector components of electric field $\mathbf{E}$, magnetic field $\mathbf{H}$, electric polarization $\mathbf{P}$, magnetization $\mathbf{m}$, and antiferromagnetic vectors $\mathbf{L}_1$, $\mathbf{L}_2$, and $\mathbf{L}_3$ for the exchange structures of bismuth ferrite, hematite, and chromite, respectively, (Fig. 2) form irreducible repre-sentations and are placed in Table 1 according to their transformation properties. For instance, it follows from this table that the $(L_z)_1$ component changes sign and the vector $L_\perp = (L_x, L_y)_1$ transforms into $(L_x, -L_y)_1$ under the symmetry operation $2_X^+$. The transformation properties of the products $M_i L_i$, $H_i E_i$, and $L_i E_i$ for bismuth ferrite are given in Table 2.

One can readily see in Table 1 that the $(H_x l_y - H_y l_x)$ combination in hematite, where $\mathbf{l} = \mathbf{L}_2/2M_0$ is the unit antiferromagnetic vector, corresponds to the first irreducible representation, i.e., is invariant. This invariant is responsible for the formation of weak magnetization $(M_x, M_y) \sim (l_y, -l_x)$ in hematite; i.e., the latter is a weak ferromagnet with the magnetization vector perpendicular to the antiferromagnetic vector.

It can also be shown that the spontaneous magnetization is forbidden in chromite $Cr_2O_3$. The combinations $H_j l_k$, where $\mathbf{l} = \mathbf{L}_3/2M_0$, are not invariant. At the same time, the chromite symmetry allows the magnetoelectric effect because of the presence of the $(E_x H_x + E_y H_y)l_z$, $E_z H_z l_z$, $H_z(E_x l_x + E_y l_y)$, and $E_z(H_x l_x + H_y l_y)$ invariants.

From Tables 1 and 2, one can easily find the following invariant for bismuth ferrite:

$$2M_0 P_z(m_y l_x - m_x l_y), \quad (1)$$

where $P_z$ is the component of spontaneous polarization $\mathbf{P} = (0, 0, P_z)$ along the c axis, $M_0$ is the magnitude of the magnetization vector of sublattices, and $\mathbf{m} = (\mathbf{M}_1 +$ $\mathbf{M}_2)/2M_0$ and $\mathbf{l} = \mathbf{L}_1/2M_0$ are the unit magnetization and antiferromagnetic vectors, respectively. This interaction has the same form as the Dzyaloshinski–Moriya interaction and, hence, also gives rise to a weak ferromagnetism with the magnetization

$$\mathbf{m} \sim (P_z l_y - P_z l_x, 0). \quad (2)$$

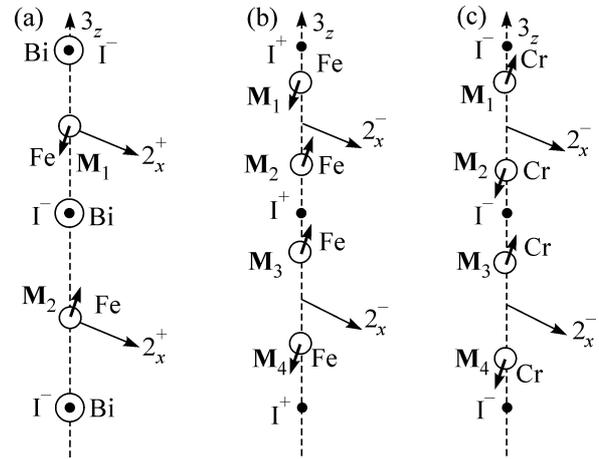

**Fig. 2.** Exchange structures of (a) bismuth ferrite $BiFeO_3$, $\mathbf{L}_1 = \mathbf{M}_1 - \mathbf{M}_2$; (b) hematite $\mathbf{L}_2 = \mathbf{M}_1 - \mathbf{M}_2 - \mathbf{M}_3 - \mathbf{M}_4$; and (c) chromite $Cr_2O_5$, $\mathbf{L}_3 = \mathbf{M}_1 - \mathbf{M}_2 + \mathbf{M}_3 - \mathbf{M}_4$.

This is by no means contradictory to the well-known theorem in the theory of antiferromagnetism about the impossibility of a weak ferromagnetism coexisting

**Table 1.** Irreducible representations of the space group $R\bar{3}c$

| | $E^+$ | $I^\pm$ | $3_z^+$ | $2_x^\pm$ | $E_i; P_i$ | $H_i; M_i$ | $L_i$ |
|---|---|---|---|---|---|---|---|
| $\Gamma_1$ | 1 | 1 | 1 | 1 | | | $(L_z)_2$ |
| $\Gamma_2$ | 1 | 1 | 1 | −1 | | $H_z, m_z$ | |
| $\Gamma_3$ | $\begin{pmatrix} 1 & 0 \\ 0 & 1 \end{pmatrix}$ | $\begin{pmatrix} 1 & 0 \\ 0 & 1 \end{pmatrix}$ | $R$ | $\begin{pmatrix} 1 & 0 \\ 0 & -1 \end{pmatrix}$ | | $\begin{pmatrix} H_x \\ H_y \end{pmatrix} \begin{pmatrix} m_x \\ m_y \end{pmatrix}$ | |
| $\Gamma_3'$ | $\begin{pmatrix} 1 & 0 \\ 0 & 1 \end{pmatrix}$ | $\begin{pmatrix} 1 & 0 \\ 0 & 1 \end{pmatrix}$ | $R$ | $\begin{pmatrix} -1 & 0 \\ 0 & 1 \end{pmatrix}$ | | | $\begin{pmatrix} L_x \\ L_y \end{pmatrix}_2$ |
| $\Gamma_4$ | 1 | −1 | 1 | 1 | | | $(L_z)_3$ |
| $\Gamma_5$ | 1 | −1 | 1 | −1 | $E_z, P_z, \nabla_z$ | | $(L_z)_1$ |
| $\Gamma_6$ | $\begin{pmatrix} 1 & 0 \\ 0 & 1 \end{pmatrix}$ | $\begin{pmatrix} -1 & 0 \\ 0 & -1 \end{pmatrix}$ | $R$ | $\begin{pmatrix} 1 & 0 \\ 0 & -1 \end{pmatrix}$ | $\begin{pmatrix} H_x \\ H_y \end{pmatrix}; \begin{pmatrix} P_x \\ P_y \end{pmatrix}; \begin{pmatrix} \nabla_x \\ \nabla_y \end{pmatrix}$ | | $\begin{pmatrix} L_x \\ L_y \end{pmatrix}_1$ |
| $\Gamma_6'$ | $\begin{pmatrix} 1 & 0 \\ 0 & 1 \end{pmatrix}$ | $\begin{pmatrix} -1 & 0 \\ 0 & -1 \end{pmatrix}$ | $R$ | $\begin{pmatrix} -1 & 0 \\ 0 & 1 \end{pmatrix}$ | | | $\begin{pmatrix} L_x \\ L_y \end{pmatrix}_3$ |

Note: $R$ is the matrix of rotation through an angle of 120° about the z axis (c axis in the hexagonal system).

**Table 2.** Irreducible representations for bismuth ferrite

| | $E^+$ | $I^-$ | $3_z^+$ | $2_x^\pm$ | $M_iL_i; H_iL_i; H_iE_i; L_iE_i;$ |
|---|---|---|---|---|---|
| $\Gamma_1$ | 1 | 1 | 1 | 1 | $(E_xL_x + E_yL_y)$ |
| $\Gamma_2$ | 1 | 1 | 1 | −1 | $(E_xL_y - E_yL_x)$ |
| $\Gamma_3$ | $\begin{pmatrix}1&0\\0&1\end{pmatrix}$ | $\begin{pmatrix}1&0\\0&1\end{pmatrix}$ | $R$ | $\begin{pmatrix}&\\&\end{pmatrix}$ | |
| $\Gamma_3'$ | $\begin{pmatrix}1&0\\0&1\end{pmatrix}$ | $\begin{pmatrix}1&0\\0&1\end{pmatrix}$ | $R$ | $\begin{pmatrix}&\\&\end{pmatrix}$ | |
| $\Gamma_4$ | 1 | −1 | 1 | 1 | $(m_xL_x + m_yL_y)$ |
| $\Gamma_5$ | 1 | −1 | 1 | −1 | $(m_yL_x + m_xL_y); (H_yL_x - H_xL_y); (H_yE_x - H_xE_y);$ |
| $\Gamma_6$ | $\begin{pmatrix}1&0\\0&1\end{pmatrix}$ | $\begin{pmatrix}-1&0\\0&-1\end{pmatrix}$ | $R$ | $\begin{pmatrix}1&0\\0&-1\end{pmatrix}$ | $\begin{pmatrix}H_yL_y - H_xL_x\\H_xL_y + H_yL_x\end{pmatrix}$ |
| $\Gamma_6'$ | $\begin{pmatrix}1&0\\0&1\end{pmatrix}$ | $\begin{pmatrix}-1&0\\0&-1\end{pmatrix}$ | $R$ | $\begin{pmatrix}-1&0\\0&1\end{pmatrix}$ | |

Note: $R$ is the matrix of rotation through an angle of 120° about the $z$ axis ($c$ axis in the hexagonal system).

with the magnetoelectric effect [24]. This theorem is related to those antiferromagnets whose space group contains space-inversion element (i.e., whose crystal structure is odd about the space inversion). In our case (BiFeO$_3$), we deal with ferroelectric magnets, where this requirement is not fulfilled. It is worth noting that the physical nature of a weak ferromagnetism in ferroelectric magnets is basically different from the Dzyaloshinski–Moriya case. A weak ferromagnetic moment in ferroelectric magnets results from the magnetoelectric interaction. In other words, this magnetic moment arises due to the magnetoelectric interaction in the internal effective electric field.

The expression for the free-energy density also includes the magnetoelectric terms proportional to the invariants of the $H_iE_il_i$ type:

$$f = \ldots - a_1|E_x(H_yl_y - H_xl_x) + E_y(H_xl_y + H_yl_x)] - a_2H_z(E_xl_y - E_yl_x) \quad (3)$$
$$- a_3E_z(H_yl_x - H_xl_y) - a_4l_z(H_yE_x - H_xE_y) + \ldots$$

For the tensor relating the magnetic-field-induced polarization vector to the magnetic vector in the linear magnetoelectric effect, one obtains

$$\alpha_{ij} = \begin{vmatrix} -a_1l_x & a_4l_z + a_1l_y & a_2l_y \\ a_1l_y - a_4l_z & a_1l_x & -a_2l_x \\ -a_3l_y & a_3l_x & 0 \end{vmatrix}. \quad (4)$$

Apart from the magnetoelectric (ME) effect and spontaneous magnetization, the magnetic symmetry of bismuth ferrite also allows the magnetic ordering of special (toroidal) type [25, 26]. The presence of toroidal moment is due to the term proportional to the

$$(\mathbf{T}[\mathbf{E} \times \mathbf{H}]) \quad (5)$$

invariant in the expression for the free energy. Hence it follows that the vector components of the toroidal moment are proportional to the antisymmetric part of the tensor of the linear ME effect:

$$T_i = \varepsilon_{ijk}\alpha_{jk}, \quad (6)$$

i.e.,

$$T_1 \sim \alpha_{23} - \alpha_{32}, \quad T_2 \sim \alpha_{31} - \alpha_{13}, \quad T_3 \sim \alpha_{12} - \alpha_{21}. \quad (6a)$$

By using Eqs. (6a) and (4), one can readily verify that the vector components of the toroidal moment in bismuth ferrite are proportional to the components of antiferromagnetic vector:

$$\begin{pmatrix}T_x\\T_y\end{pmatrix} \sim \begin{pmatrix}l_x\\l_y\end{pmatrix}; \quad T_z \sim l_z. \quad (7)$$

The analysis of the irreducible representations of crystal space groups allows one to predict not only the properties of magnetic materials but also their microscopic magnetic structure. Taking into account that the differential operator $\nabla$ transforms as a polar vector, we can write the new term in the expression for the free-energy density (Table 1):

$$f_L = \gamma P_z (l_x \nabla_x l_z + l_y \nabla_y l_z - l_z \nabla_x l_x - l_z \nabla_y l_y), \quad (8)$$

where $\gamma$ is the coefficient of the ME origin. Note that, by analogy with the invariant of the form $l_i(\partial l_j/\partial x) - l_j(\partial l_i/\partial x)$, combination (8) was named the Lifshitz invariant. However, the distinctive feature of combination (8) is that it is nonzero only in the presence of spontaneous polarization. The Lifshitz invariant (8) has the ME origin, but, contrary to the homogeneous ME interaction of the Dzyaloshinski–Moriya type, it is nonzero only in the presence of an inhomogeneous antiferromagnetic structure, i.e., only if the spatial derivatives of the antiferromagnetic vector are nonzero.

Therefore, two types of ME interaction are present in bismuth ferrite: a homogeneous interaction characterized by the tensor $\alpha_{ij}$ of the linear ME effect, and an inhomogeneous interaction characterized by the $\gamma$ constant. A distinctive feature of bismuth ferrite is the presence of spontaneous polarization in it. This implies that the space-inversion element is absent in its crystallographic space group and manifests itself in the coexistence of weak ferromagnetism (2), ME effect (4), and inhomogeneous ME interaction (8).

Interestingly, the inversion symmetry of chromite $Cr_2O_3$ is violated upon applying the external electric field $E^{ext}$ along the $c$ axis, and the spontaneous magnetization arises:

$$\mathbf{m} \sim (E^{ext} l_x, E^{ext} l_y, 0), \quad (9)$$

where $\mathbf{l} = \mathbf{L}_3/2M_0$ is the unit antiferromagnetic vector.

## IV. SPATIALLY MODULATED STRUCTURE AND MAGNETIC-INDUCED PHASE TRANSITIONS

The total expression for the free-energy density is has the form

$$f = f_L + f_{exch} + f_{an}, \quad (10)$$

where

$$f_{exch} = A \sum_{i=x,y,z} (\nabla l_i)^2 = A((\nabla \theta)^2 + \sin^2\theta (\nabla \varphi)^2) \quad (11)$$

is the exchange energy; $A = 3 \times 10^7$ erg/cm is the constant of inhomogeneous exchange (exchange stiffness);

$$f_{an} = K_u \sin^2\theta \quad (12)$$

is the anisotropy energy; $\theta$ and $\varphi$ are the polar and azimuthal angles of the unit antiferromagnetic vector; and $\mathbf{l} = (\sin\theta\cos\varphi, \sin\theta\sin\varphi, \cos\theta)$ in the spherical coordinate system with the polar axis aligned with the principal axis $c$ (hexagonal system).

The minimization of the free-energy functional $F = \int f\, dV$ by the Lagrange–Euler method in the approximation ignoring anisotropy gives for the functions $\theta(x, y, z)$ and $\varphi(x, y, z)$ [27, 28]:

$$\varphi_0 = const = \arg\tan(q_y/q_x), \quad \theta_0 = q_x x + q_y y, \quad (13)$$

where $\mathbf{q}$ is the spiral wave vector. Solution (13) describes the cycloid whose plane is perpendicular to the basal plane and oriented along the propagation direction of the modulation wave.

A more precise solution allowing for the anisotropy $K_u$ of the material is given by the formula [28, 29]

$$\cos\theta = sn(q_x x, m = -K_u/E), \quad (14)$$

which corresponds to an anharmonic cycloid. For the anisotropy constant much smaller than the exchange energy, $K_u \ll Aq^2$, the module $m$ of elliptic sine tends to zero, and solution (14) turns to harmonic solution (13).

It is worth noting that the presence of spatially modulated structure was confirmed not only by neutron diffraction but also by nuclear magnetic resonance [21–23]. Instead of a single peak corresponding to a homogeneous structure, the NMR line was more complicated and showed two maxima corresponding to the spin orientations perpendicular and parallel to the principal axis (Fig. 3). The analysis of the NMR lineshape not only showed the presence of a cycloid but also allowed the spin distribution to be reconstructed along the cycloid. At low temperatures ($T = 4.2$ K), this distribution proved to be essentially anharmonic: for the most part of cycloid period, spins form a small angle with the axis, as was seen from the higher intensity of the high-frequency peak. With an increase in temperature, the lineshape becomes more symmetric, the anharmonicity decreases, and, at room temperature, the coordinate dependence of the angle becomes close to linear [21] (Figs. 3a, 3b). By substituting Eq. (13) in Eq. (10), one obtains for the volume-averaged free-energy density in the approximation of harmonic cycloid

$$\langle F \rangle = Aq^2 - (\gamma P_z)q + K_u/2. \quad (15)$$

The wave vector corresponding to the energy minimum is

$$q = \frac{2\pi}{\lambda} = \frac{\gamma P_z}{2A}. \quad (16)$$

Knowing the structure period ($\lambda = 620$ Å) and assuming that the polarization is $P_z = 6 \times 10^{-6}$ C/cm² mol and the exchange constant is $A = 3 \times 10^{-7}$ erg/cm, one can estimate the inhomogeneous ME coefficient $\gamma = 10^5$ erg/C = $10^{-2}$ V.

Because of the presence of cycloid, the volume-averaged ME effect (4), the spontaneous magnetization (2), and the toroidal moment (7) are zero. All three effects appear only upon the suppression of the spatially modulated structure.

The modulated structure can be suppressed, e.g., by applying a strong magnetic field. Magnetic field adds the following term to the expression for the anisotropy constant (field is applied along the threefold axis $c$):

$$K_u = K_u^0 - \chi_\perp \frac{H_z^2}{2}, \quad (17)$$

where $\chi_\perp$ is the magnetic susceptibility in the direction perpendicular to the antiferromagnetic vector, and $K_u^0$ is the uniaxial anisotropy constant in zero field. For the fields higher than a certain critical value, the spatially modulated structure may become energetically unfavorable, as compared to the homogeneous state whose energy is given by the formula

$$E_{\text{hom}} = K_u, \quad (18)$$

where $K_u$ is effective anisotropy (17).

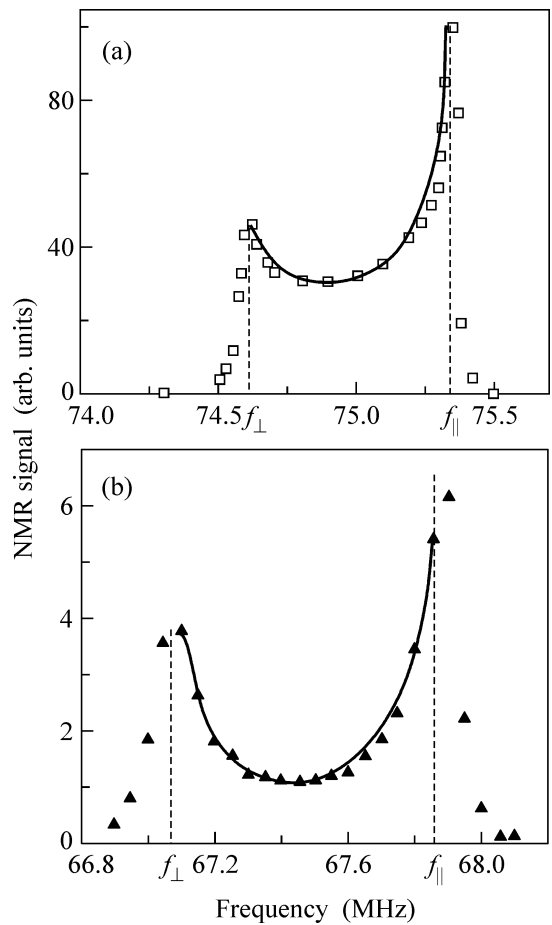

**Fig. 3.** $^{57}$Fe NMR spectra of BiFeO$_3$ at (a) 77 and (b) 304 K [21].

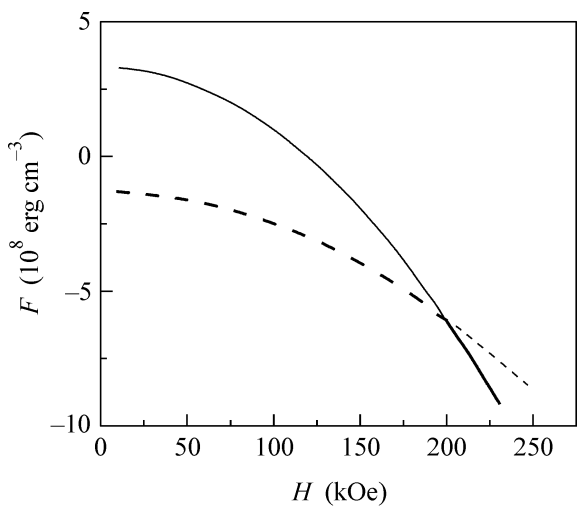

**Fig. 4.** Dependence of the free-energy density on the magnetic field. The solid line is for the homogeneous state and the dashed line is for the spatially modulated state.

The field dependences of the free-energy density are presented in Fig. 4 for spatially modulated structure (15) and homogeneous state (18). Although the spatially modulated structure at low fields is energetically more favorable than the homogeneous state, the situation changes in the range of high fields. The critical field $H_c$, at which the phase transition occurs, is determined from Eqs. (15) and (18):

$$H_c = \sqrt{\frac{2(K_u^0 + 2Aq^2)}{\chi_\perp}}. \quad (19)$$

The estimate with $K_u^0 \ll Aq^2$, $A = 3 \times 10^{-7}$ erg/cm, and $\chi_\perp = 4.7 \times 10^{-5}$ gives for the critical field ~200 kOe. The phase transitions in bismuth ferrite with an anharmonic cycloid were considered theoretically in [29].

The suppression of the spatially modulated structure in bismuth ferrite and appearance of the linear ME effect and the toroidal moment were confirmed experimentally by measuring the field dependence of polarization in pulsed fields [27, 30, 31]. At $H < H_c$, the polarization depends on field almost quadratically, whereas, at the critical field, the polarization undergoes an abrupt jump that is accompanied by the appearance of the linear ME effect and the renormalization of the quadratic ME tensor (Fig. 5). The values of the critical field and the linear ME effect were found to be ~200 kOe and ~$10^{-10}$ C/(m$^2$ Oe), respectively.

The orientation of the cycloid plane in crystal was determined from the measurements of electric $a$ and $b$ polarizations as functions of a magnetic field applied along the crystal $c$ axis [31]. The $P_a(H_c)$ and $P_b(H_c)$ dependences are shown in Fig. 6. One can see that, after the cycloid disappears in the field $H_c \approx 200$ kOe, the electric polarization increases jumpwise and varies at $H > 250$ kOe almost linearly with field. It would be natural to assume that the spins of Fe$^{3+}$ ions do not offset from the cycloid plane in the resulting structure but "lie down" in the basal plane so that $L_z = 0$ and $L_y/L_x = \tan\varphi$, where $\varphi$, according to the neutron diffraction data, is the angle between the cycloid plane and the crystal $a$ axis. Then, in the linear approximation (Eq. (4)), one obtains from the slopes of the linear portions of experimental curves:

$$\frac{dP_a/dH}{dP_b/dH} = \frac{\alpha_{13}}{\alpha_{23}} = -\frac{l_y}{l_x} = -1.8. \qquad (20)$$

The resulting value $\varphi = -60°$ corresponds to the angle obtained in [20] from the neutron diffraction data ($[110]_h$ direction in Fig. 1b).

In [27, 30, 31], measurements were carried out in pulsed fields. Recent experiments [32] on the observation of antiferromagnetic resonance in a static magnetic field have shown that the eigenfrequency spectrum of bismuth ferrite strongly changes near the point $H_c = 180$ kOe (Fig. 7). The authors of that work assigned this change to the phase transition from the spatially modulated structure to the homogeneous state [32]. It was also shown that the phase transition is accompanied by an appreciable hysteresis of the resonance lines in the increasing and decreasing fields (Fig. 8). A value of 3.3 V/(cm Oe) found for the ME coefficient from the curve-fitting procedure in the above- critical fields is in a good agreement with the giant ME effect in thin bismuth ferrite films.

## V. WEAK FERROMAGNETISM

The symmetry analysis of crystal properties suggests that a weak ferromagnetism caused by the canted sublattice magnetizations $\mathbf{M}_1$ and $\mathbf{M}_2$ may occur in bismuth ferrite.

Spontaneous magnetization (2) depends on the mutual arrangement of the polarization and antiferromagnetic vectors and is proportional to the sine of the angle between the polarization and antiferromagnetic vectors:

$$m \sim PL\sin\theta. \qquad (21)$$

It follows from Eq. (21) that the canting angle $\varphi$ between the sublattice magnetizations $\mathbf{M}_1$ and $\mathbf{M}_2$ is determined as

$$\varphi = m/L \sim \sin\theta, \qquad (22)$$

i.e., it is maximal if the antiferromagnetic vector is perpendicular to the crystal principal axis and is zero if it is parallel to the latter.

The presence of the spatially modulated structure results in a periodic variation of the canting angle between the sublattices, so that magnetization (21) averaged over the cycloid period is zero.

It is natural to expect that the spontaneous magnetization can arise in the fields higher than the critical field $H_c$, in which the spatially modulated structure disappears and a homogeneous antiferromagnetic ordering is established.

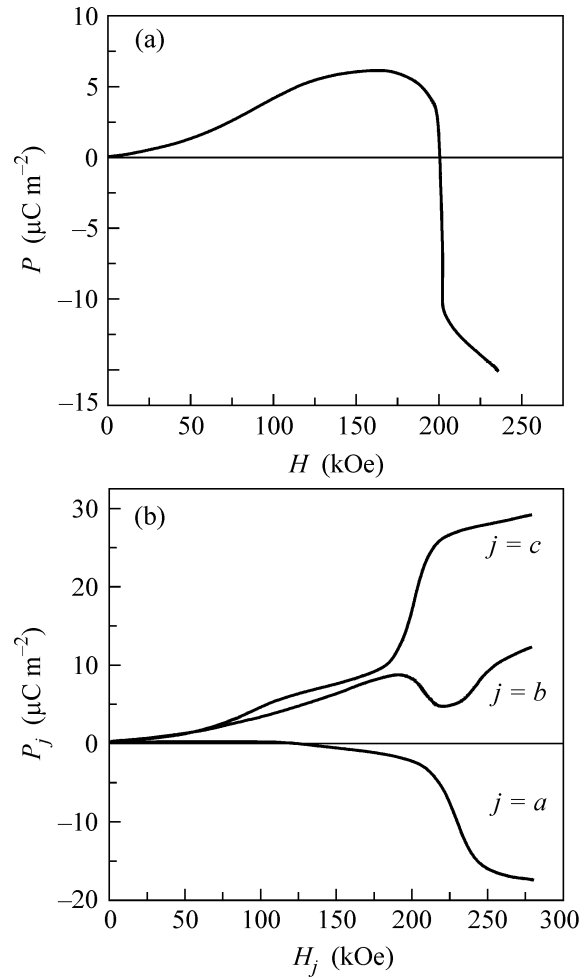

**Fig. 5.** Dependence of the longitudinal electric polarization on the magnetic field for different crystallographic directions: (a) along the cubic axis and (b) along the $a$, $b$, and $c$ axes in the hexagonal system [27, 30].

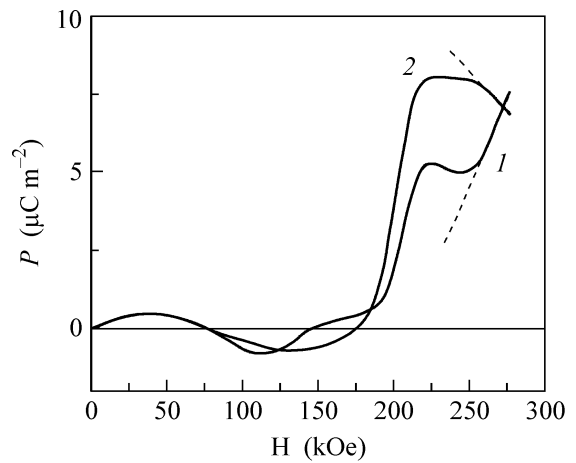

**Fig. 6.** Experimental curves obtained for the polarizations along different directions in a magnetic field $H \parallel c$ at a temperature of 18 K: (*1*) along the $b$ axis and (*2*) along the $a$ axis. The dashed lines show the slopes of the linear portions of the curves at $H > 250$ kOe, which were used to calculate the derivatives $dP_a/dH$ and $dP_b/dH$ by formula (20).

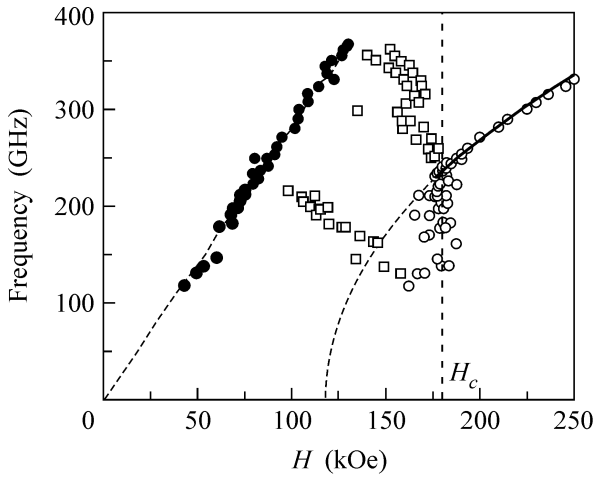

**Fig. 7.** Antiferromagnetic resonance frequencies as functions of magnetic field $H$ ($T = 4.2$ K) [32].

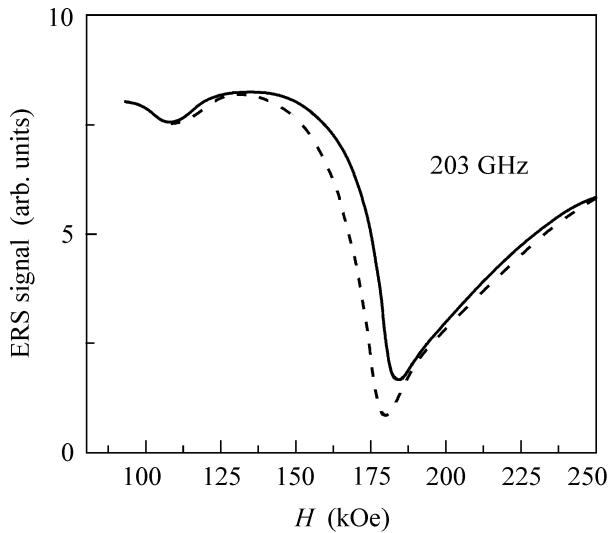

**Fig. 8.** Magnetic hysteresis of the absorption peak. Solid line is for the increasing field and dashes are for the decreasing field.

In the very recent past [34], the magnetization of BiFeO$_3$ has been measured as a function of magnetic field (Fig. 9). Measurements were made at $T = 10$ K for the $[001]_c$ direction (along the edge of a crystal with natural faces shaped almost like the cube faces). The identical magnetization curves were also obtained for the $[100]_c$ and $[010]_c$ directions. Measurements for the **H** ∥ **c** geometry were not carried out because of a poor sensitivity of the setup (this could be done only with small-sized oriented samples).

It is seen that the run of magnetization curve changes sharply at the fields close to critical $H_c \sim$ 200 kOe because of the suppression of cycloid by the magnetic field and transition to a homogeneous antiferromagnetic phase.

The experimental magnetization curves at low and high fields are well described by the linear dependences

$$M = \chi H, \quad H < 100 \text{ E}, \tag{23a}$$

$$M = M^{\text{spont}}_{[001]} + \chi_\perp H, \quad H > H_c, \tag{23b}$$

i.e., the spontaneous magnetization $M^{\text{spont}}_{[001]} = 0.25 \pm 0.02$ (G cm$^3$)/g (1.95 $\pm$ 0.15 G) arises in bismuth ferrite above the critical field $H_c$, while the field dependence in this range is described by a linear function with the slope equal to the susceptibility $\chi_\perp = 0.6 \times 10^{-5}$ (G cm$^3$)/(g Oe) of the material in the direction perpendicular to the antiferromagnetic vector [34]. In low fields ($H < 100$ kOe), the field dependence of magnetization is nicely described by the direct proportionality, where the susceptibility $\chi$ comprises approximately 5/6 of the total susceptibility $\chi_\perp$. In the range of intermediate fields (100 kOe $< H <$ 200 kOe), the dependence is noticeably nonlinear.

The observed experimental dependence can be understood if the spontaneous magnetization $\langle M^{sp} \rangle$ averaged over the cycloid period and the magnetic-induced magnetization $M^H$ depend on the external field **H**:

$$\mathbf{M}(H) = \langle \mathbf{M}^{\text{spont}} \rangle (H) + \mathbf{M}^H(H). \tag{24}$$

Considering that the field in the experiment is aligned with the $[001]_c$ axis, i.e., $\mathbf{H} = (H\sqrt{2}/\sqrt{3}, 0, H/\sqrt{3})$, one has for the total magnetization along the $[001]_c$ direction

$$M_{[001]}(H) = M^{\text{spont}}_{[001]} \langle \sin\theta \rangle_\lambda + \underbrace{\frac{2}{3}\chi_\perp H}_{M^H_{x[001]}} + \underbrace{\frac{1}{3}\chi_\perp H \langle \sin^2\theta \rangle_\lambda}_{M^H_{z[001]}}, \tag{25}$$

where $\theta$ is the angle between the spontaneous magnetization $P_z$ ($c$ axis) and the antiferromagnetic vector and $M^H_{x[001]}$ and $M^H_{z[001]}$ are, respectively, the projections of the $M_x$ and $M_z$ components onto the $[001]_c$ direction (Fig. 10). Equation (25) can be used to explain the experimentally observed dependences (23). The values of sine $\langle \sin\theta(y) \rangle_\lambda$ and the square of sine $\langle \sin^2\theta(y) \rangle_\lambda$ averaged over the spatial period are different for the low $H \longrightarrow 0$ and high $H \longrightarrow H_c$ fields and vary within (0, 1) and (1/2, 1), respectively. Dependence (25) turns to (23a) at $H \longrightarrow 0$ and to (23b) at $H \longrightarrow H_c$.

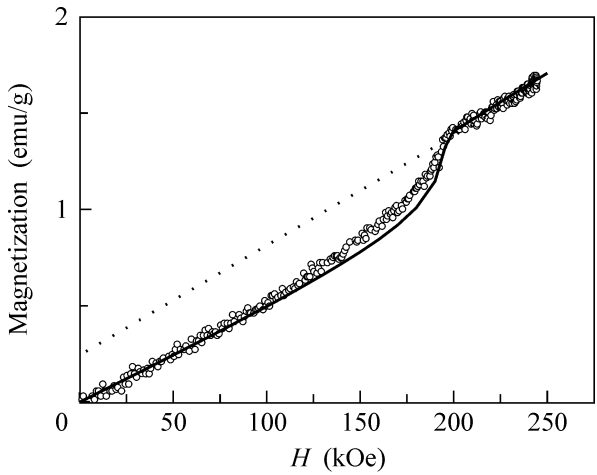

**Fig. 9.** Magnetization of bismuth ferrite as a function of magnetic field at a temperature of 10 K [34]. Dots are the experimental data obtained in a field oriented along the $[001]_c$ direction, and solid line is the theoretical dependence (Eq. (28)).

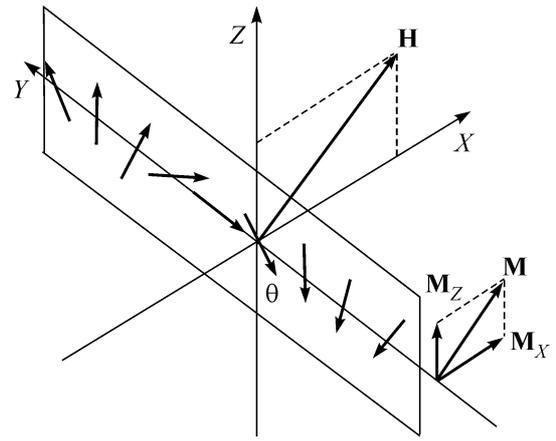

**Fig. 10.** Cycloidal spin structure in an external magnetic field. The directions of the field **H** and magnetization **M** are different because the spatially modulated spin structure is anisotropic.

## VI. TOROIDAL ORDERING

Measurements of the field dependence of electric polarization showed that a toroidal moment appears upon the magnetic phase transition in bismuth ferrite [31].

To determine the toroidal moment in $BiFeO_3$, the antisymmetric part (proportional to the toroidal moment) of the tensor of the linear ME effect was examined at fields exceeding the phase-transition field $H_c$ and oriented at an angle of 45° to the crystal $a$ and $b$ axes in the basal plane. Such a field orientation made it possible to measure the polarization components $P_a(H)$ and $P_b(H)$ in the same state of the sample.

For the indicated field orientation and $H > H_c$, the antiferromagnetic vector is aligned with the threefold axis $L \| c$ ($L_x = L_y = 0$) and the crystal belongs to the 3m magnetic class, so that $\alpha_{12}$ and $\alpha_{21}$ (see Eq. (4)) are the only nonzero components, and, in addition,

$$\alpha_{12} = -\alpha_{21}. \quad (26)$$

The matrix $\beta_{ijk}$ of the ME effect quadratic in magnetic field is more cumbersome, and the nonzero matrix elements in this case are

$$\beta_{111} = \gamma_1, \quad \beta_{122} = \beta_{212} = \beta_{221} = -\gamma. \quad (27)$$

Thus, taking into account that $H_a = H_b = H/\sqrt{2}$, one finds

$$P_a = \frac{\alpha_{12}}{\sqrt{2}} H, \quad P_b = \frac{\alpha_{21}}{\sqrt{2}} H - \gamma_1 H^2. \quad (28)$$

By fitting the tails of experimental curves (Fig. 11) to the linear (for $P_a(H)$) and quadratic (for $P_b(H)$) approximations (the approximating straight line and parabola are shown by dashes), the following ME coefficients were obtained:

$$\alpha_{12} = -(0.029 \pm 0.003) \times 10^{-6} \, \text{Kl/(m}^2 \, \text{kE)},$$

$$\alpha_{21} = +(0.032 \pm 0.003) \times 10^{-6} \, \text{Kl/(m}^2 \, \text{kE)}, \quad (29)$$

$$\gamma_1 = 5 \times 10^{-11} \, \text{Kl/(m}^2 \, \text{kE)},$$

Thus, the obtained $\alpha_{12}$ and $\alpha_{21}$ components differ only in sign and, to within the error, are equal in magnitude, in full agreement with Eq. (29). The fact that the nondiagonal tensor components of the linear ME effect are antisymmetric is evidence that the toroidal moment $T_z \sim (\alpha_{12} - \alpha_{21})$ will appear after the suppression of cycloid.

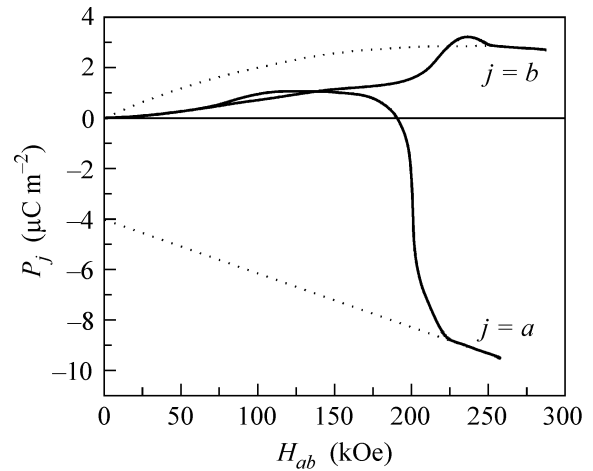

**Fig. 11.** The polarizations along the $a$ and $b$ directions as functions of a field applied in the basal plane at an angle of 45° to the $a$ axis [31]. Dotted curves are the theoretical dependences calculated by Eqs. (28).

## VII. NEW MATERIALS WITH GIANT MAGNETOELECTRIC EFFECT

Apart from the strong magnetic field, other methods exist of suppressing the spatially modulated structure. One such method is the substitution of rare-earth ions for the bismuth ions. The compounds with formula $RFeO_3$ (rare-earth orthoferrites) also have the perovskite structure, though orthorhombically distorted. The introduction of rare-earth impurities can increase the anisotropy constant to an extent that the presence of spatially modulated structure would be energetically unfavorable. Measurements of the magnetic-field dependences of polarization in the $Bi_{1-x}R_xFeO_3$ (R = La, Gd, Dy; $0.4 < x < 0.5$) compounds [35–38] showed the presence of the linear ME effect up to the liquid nitrogen temperature, evidencing the suppression of the spatially modulated spin structure in the compounds of this type. The high-field polarization measurements showed that the spatially modulated structure is retained in bismuth ferrite with a smaller lanthanum content ($x < 0.3$), but the presence of lanthanum additives reduces the transition field from the spatially modulated to the homogeneous state [37, 38].

Interesting results were obtained in thin bismuth ferrite films of size 50–500 nm [33], where the observed ME effect $dE/dH = 3$ v/(cm Oe) was an order of magnitude greater than in the classical chromite magnetoelectric. According to the presently accepted nomenclature, such effects are referred to as giant effects. The modulated structure in these compounds is suppressed likely due to the stresses that arise during the epitaxial growth and can produce, through the magnetostriction or piezoelectric mechanism, the magnetic and electric fields that are critical to the phase transitions.

In a search for the materials exhibiting the giant ME effect, attention of researchers has also be drawn to the bismuth ferrite-based solid solutions $BiFeO_3$–$xPbTiO_3$ with the substitution of lanthanum for the bismuth ions [39]. In these compounds, an anomalously high electric polarization and the appearance of spontaneous magnetization are observed near the point of phase transition ($PbTiO_3$ content $x \sim 30\%$ [40]) from the orthorhombic to the tetragonal phase [39]. These two phenomena can serve as an indirect evidence of the presence of a rather strong ME effect in these materials.

## VIII. CONCLUSIONS

Symmetry analysis has been applied to the magnetic and magnetoelectric properties of antiferromagnets. It has been shown that the violation of the inversion symmetry in both crystal and magnetic structures of the bismuth ferrite antiferromagnet $BiFeO_3$ gives rise to the unique possibility of the linear ME effect coexisting with a weak ferromagnetic moment in $BiFeO_3$, which is basically impossible for conventional antiferromagnets. The polarization breaks the crystal central symmetry and manifests itself also in the appearance of an inhomogeneous ME interaction (Lifshitz invariant), leading to the formation of a spatially modulated spin structure in the material. The suppression of the spatially modulated spin structure is the necessary condition for the presence of the linear ME effect, the spontaneous magnetization, and the toroidal moment. One of the ways of suppressing the spin cycloid consists in applying a strong magnetic field, which induces the phase transition from the spatially modulated state to the homogeneous state, after which all three above effects appear. Other methods of suppressing the spin cycloid (the substitution of rare-earth ions for the bismuth ions and the fabrication of bismuth ferrite epitaxial films) made it possible to obtain a giant ME effect that is an order of magnitude greater than the previously observed effect.